# Interactions Between Bilayers of Phospholipids Mixture Extracted from Human Osteoarthritic Synovial Fluid

Yifeng Cao [a], Nir Kampf [a], Marta Krystyna Kosinska [b], Juergen Steinmeyer [b], and Jacob Klein [a]*

a. Department of Materials and Interfaces, Weizmann Institute of Science, Rehovot 76100, Israel

b. Laboratory for Experimental Orthopaedics, Department of Orthopaedics, Justus-Liebig-University Giessen, 35392 Giessen, Germany

*To whom correspondence should be addressed: Jacob.Klein@weizmann.ac.il

**Abstract**

Duncan Dowson, whom this issue commemorates, was a world leader in the field of biotribology, with prolific contributions both in fluid-based and boundary lubrication of biological tissues, in particular articular cartilage, a central issue in biotribology due to its importance for joint homeostasis. Here we explore further the issue of cartilage boundary lubrication, which has been attributed to phospholipid (PL)-exposing layers at the cartilage surface in part. A surface force balance (SFB) with unique sensitivity is used to investigate the normal and frictional interactions of the boundary layers formed by PLs extracted from osteoarthritic (OA) human synovial fluid (hSF). Our results reveal that vesicles of the OA-hSF lipids rupture spontaneously to form bilayers on the mica substrate (which, like the *in-vivo* articular cartilage surface in synovial joints, is negatively-charged) which then undergo hemifusion at quite low pressures in the SFB, attributed to the large heterogeneity of the hSF lipids. Nanometric friction measurements reveal friction coefficients $\mu \approx 0.03$ across the hemi-fused bilayer of these lipids, indicating residual hydration lubrication at the lipid-headgroup/substrate interface. Addition of calcium ions causes an increase in friction to $\mu \approx 0.2$, attributed either to calcium-bridging attraction of lipid headgroups to the negatively-charged substrate, or a shift of the slip plane to the more dissipative hydrophobic-tail/hydrophobic-tail interface. Our results suggest that the heterogeneity and composition of the OA-hSF lipids may be associated with higher friction at the cartilage boundary layers – and thus a connection with greater wear and degradation – due to hemifusion of the exposed lipid bilayers.

**Keywords**

Biolubrication; articular-cartilage; lipid-based lubrication; synovial fluid lipids; surface forces; hemifusion.

# 1. Introduction

Professor Duncan Dowson, who passed away earlier this year (2020) and whose memory this issue of *Biotribology* celebrates, was one of the leaders world-wide in the field of tribology. He was one of the last surviving members of – and one of only two academics in – the celebrated Jost Commission [1], which in 1966 issued a seminal report on the economic burden of friction and wear, and indeed coined the word 'tribology'. Much of Duncan Dowson's prolific scientific activity centered in the area of friction and lubrication in biological systems [2], where he had influential contributions, so that this memorial issue is particularly appropriate. One of us (JK) especially recalls asking him once concerning the 'true' boundary friction of articular cartilage *in vivo*, given that any measurements on friction between articular cartilage surfaces in living joints must also be affected by viscous dissipation arising from the distortion of adjacent tissues and ligaments as the joint articulates. Duncan's response revealed that he had anticipated this issue long before, and referred to (unpublished) experiments trying to eliminate precisely these extraneous viscous losses, by excision of adjacent tissues. To this day that insight is highly relevant, and is cited by our group as a valued 'private communication' when referring to the difficulties of measuring the very low boundary friction of articular cartilage *in vivo*.

Indeed, the extremely low friction of healthy articular cartilage in synovial joints, providing life-long efficient lubrication (with measured friction coefficients of $10^{-2}$ – $10^{-3}$ or lower – see above D. Dowson – private communication) and wear protection, while bearing high loads and experiencing local contact pressures up to over 5 MPa [[3], [4], [5], [6]], is essential for their homeostasis. It is generally accepted that joint lubrication is enabled by both fluid-based mechanisms (such as weeping or interstitial fluid pressurization [[7], [8]] and boundary lubrication [[2], [9], [10]]. In the boundary lubrication mode, friction is regulated by a molecularly-thin layer coating the outer surface, i.e. the boundary, of each of the opposing articular cartilage layers. Synovial fluid, which plays a key role in the functions of synovial joints, acts also to maintain and replenish these boundary layers [[11], [12]].

There have been extensive efforts to identify and characterize the components of the cartilage boundary layer, especially those also present in synovial fluid, as these would shed light on the joint lubrication mechanism at a molecular level [[13], [14], [15], [16], [17]]. Such identification may enable more efficient treatment of articular cartilage diseases, particularly osteoarthritis (OA), a common degenerative joint disorder arising from cartilage damage, affecting millions worldwide [18]. According to recent proposals [[10], [12], [19]], cartilage boundary layers expose at their outer surface (i.e. at the slip plane) phosphatidylcholine (PC) lipids that are complexed with other joint macromolecules including hyaluronan (hyaluronic acid or HA) and lubricin, acting together; so that removing one of these components may lead to increase in friction [[12], [20], [21]]. A schematic of this proposed boundary layer is shown in Fig. 1.

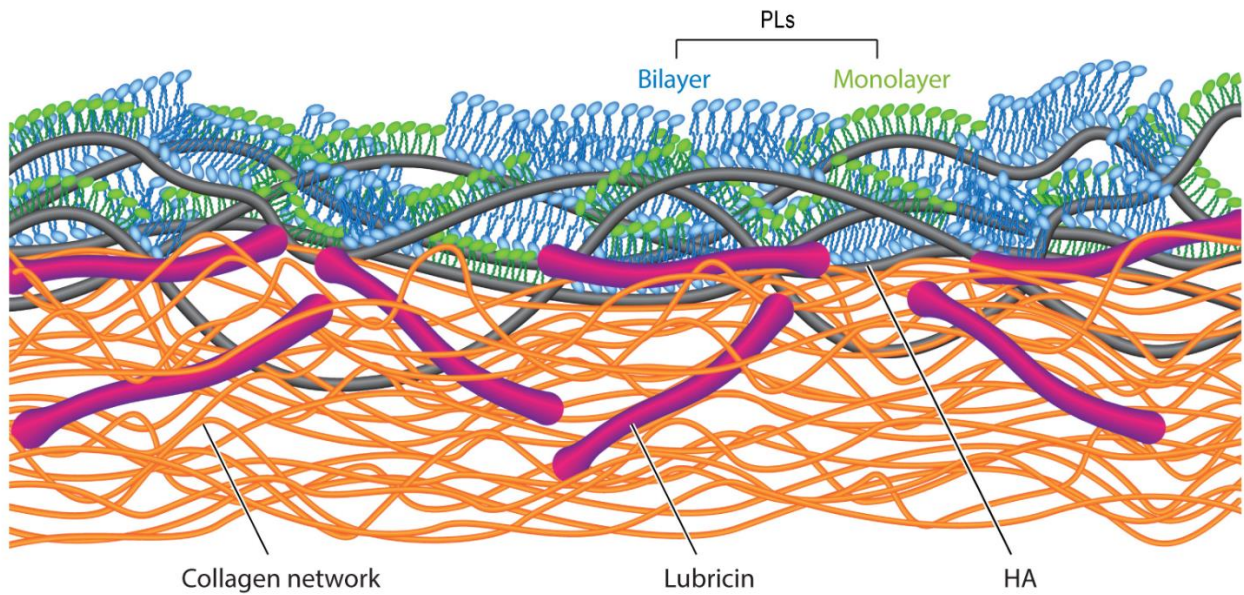

**Fig. 1**. Illustrating the proposed boundary layer on articular cartilage, where lubricin molecules anchor HA chains at the articular cartilage surface and the HA is complexed with PL layers exposing their hydrated headgroups at the very outer surface (the slip plane). Adapted from Ref. [10].

HA and proteins (such as glycosaminoglycans, which are ubiquitous in cartilage) do not by themselves lubricate efficiently at the high contact pressures found in joints [[10], [22], [23]]. For the case of lubricin on its own, a number of studies have indicated high friction [[12], [23], [24]], though a more recent one suggested much lower values [25]. PC lipids, the most abundant PLs in joints, may however provide efficient boundary lubrication and wear protection up to physiologically high pressures [[12], [25], [26], [27], [28], [29]]. This is attributed to hydration lubrication mediated by their highly hydrated phosphocholine headgroup layers exposed by the cartilage boundary layer [[10], [12], [19], [30], [31]]. It should be emphasized that all model studies to date on lubrication by PL-based boundary layers have involved single component lipids (or at most a binary PL mixture [32]) rather than the large numbers of different lipids and lipid types that are present in living joints [[33], [34], [35], [36]].

The efficiency of lubrication by PL layers depends strongly on which lipids are exposed at the slip plane, and this in turn must be related to the actual composition and concentration of lipids in the surrounding SF, since the robustness of the PL layers and their propensity for self-healing is a function of these [[30], [32]]. It has been found that in OA joints the total PL content and the composition of the lipids themselves in the SF differ from that in healthy joints [[35], [36]]. Since the OA-associated cartilage degradation with wear is due to increased boundary friction, it is thus of interest to examine boundary lubrication using lipids extracted from SF in OA-afflicted joints. Contradictory results have been reported for the relationship between the friction coefficient in healthy and in osteoarthritic joints [[37], [38], [39], [40]]; these however have come from macroscopic tribometry studies which cannot directly provide information on interactions at a molecular level.

To get better insight into this question, therefore, we examine in this study the normal and shear forces between layers composed of lipids directly extracted from OA human synovial fluid (hSF). We do this using the surface force balance (SFB) technique, which is capable of sub-nanometer resolution of surface separation and of uniquely-sensitive force measurements. Due to the negatively charged characteristic of PLs mixture from hSF and the role of divalent cations, largely calcium, in interactions between anionic PLs and negatively charged biomacromolecules identified in synovial joints [41], we also examine here the effect of calcium ions (at their level in synovial fluid) on the interactions between such PL-based boundary layers [42].

## 2. Material and methods

### 2.1 Materials

Sodium nitrate (NaNO$_3$, 99.99 Suprapur®, Merck) and calcium nitrate tetrahydrate (Ca(NO$_3$)$_2$·4H$_2$O, 99.95 Suprapur®, Merck) were used as received. Conductivity water with a resistivity of 18.20 MΩ·cm at room temperature and total organic carbon ⩽ 2 ppb was obtained using a Thermo Fisher Barnstead Nanopure water purification system.

### 2.2 Liposomes preparation

PLs from hSF of 17 patients (11M, 6F; age range 25 – 49) diagnosed with OA at a similarly-early stage of OA development were extracted and characterized [[35], [36]]; the mean concentrations of the different PL classes were determined (and appear in Fig. 2) and did not depend significantly on gender, BMI or age, though their concentrations differed significantly from corresponding PLs extracted from healthy hSF [[35], [36]]. The sample used in our study, from hSF of a patient (M; age 41) in a similar OA stage to these 17 patients, was within the range spread of their mean values (see Fig. 2) and may thus be taken as representative of PLs from hSF at this OA stage. The procedure was approved by the local ethics committee of Justus-Liebig-University of Giessen, and written informed consent was obtained from all patients. Procedures for synovial fluid sampling and PLs extraction have been elaborated in details in previous studies [35]. Briefly, after sampling, proteinase- and phospholipase-inhibitors (PPIs), including quinacrine·2HCl, neomycin sulfate, gentamycin sulfate, and proteinase inhibitor (Roche, Cat. NO. 11697498001), were added to synovial fluid which was frozen at -86 °C until PLs were extracted. PLs were extracted from hyaluronidase-treated synovial fluid, dried by blowing nitrogen, kept in dry ice during delivery and at -20 °C before use. Liposomes at a total PL concentration of ca. 0.3 mM were prepared from PLs extracted from OA-hSF, by adding 150 mM $NaNO_3$ solution to the dried PLs. The mixture was then sonicated at 65 °C and vortexed vigorously to form multilamellar vesicles. Small unilamellar vesicles (SUVs) were prepared by successively extruding the MLVs through three polycarbonate membranes (Whatman, GE Healthcare, USA) with pore sizes of 400, 100, and 50 nm for 5, 8, and 12 times, respectively, with an extruder (Northern Lipids, Burnaby, Canada) kept at 65 ± 1 °C. The prepared dispersion of liposomes, designated hSF-PLs-SUVs (human synovial fluid phospholipids small unilamellar vesicles) was kept at 4 °C before use.

**2.3 Size distribution and zeta potential**

The size distribution and zeta potential of the prepared liposomes were determined by dynamic light scattering (DLS) technique (Malvern Zetasizer Nano ZSP instrument), using a backscatter angle of 173°. Folded capillary cells (DTS1070) were used for both DLS and zeta potential measurements. For zeta potential, the Smoluchowski model was selected. Temperature was set at 25 °C for all the measurements.

The liposome size and concentration were measured by nanoparticle tracking analysis (NTA) (Malvern NanoSight NS300). Before the liposome sample was injected to the cell, it was checked with pure water to ensure it was clean. Scattered light from the particles was captured by a CCD camera. For each measurement, a 60 s video at a frame rate of 24 frames/second was taken for three times at different positions in the measuring cell. Three measurements were carried out for each sample. Data analysis was performed using the NanoSight NTA software.

For both DLS and NTA method, the hydrodynamic diameter of the particle ($d$) was calculated according to Stokes-Einstein equation, $d = K_b T/3\pi\eta D_{diff}$, where $K_b$ is the Boltzmann's constant, $T$ is the absolute temperature, $\eta$ is viscosity, and $D_{diff}$ is the diffusion coefficient. In DLS, $D_{diff}$ is measured by time dependent scattering intensity fluctuations with a digital correlator; while in NTA, $D_{diff}$ is calculated by tracking the individual particle diffusive motion with video.

## 2.4 Atomic force microscopy (AFM)

Morphology of the adsorbed liposomes on mica surface was revealed by AFM (MFP-3D SA, Oxford Instruments Asylum Research, Inc., Santa Barbara), using a Silicon tip (spring constant 0.35 N/m) on a Silicon Nitride lever (SNL-10, Bruker Nano Inc, USA), in AC mode under aqueous conditions. For liposome adsorption on mica, samples were prepared by incubating freshly cleaved mica in ca. 0.3 mM liposome dispersion for more than 4 hours without passing air/water interfaces. For the surface scanned after the SFB experiment, the sample perforce passed the air/water interface twice while transferring from SFB setup to AFM sample holder and scanning was performed under water.

## 2.5 SFB

Normal and shear forces between OA-hSF-PLs adsorbed on mica were measured by an SFB (schematic inset in Fig. 5(a) below), as described in detailed elsewhere [43]. In brief, two single crystallographic mica facets were back-silvered and glued on cylindrical quartz lenses in a crossed-cylinder configuration. Separation distance $D$ between the two surfaces, and their mean radius of curvature $R$, were determined by a multiple-beam interferometric technique via the wavelengths and shapes of fringes of equal chromatic order (FECO). Normal and shear forces ($F_n$ and $F_s$) were determined by monitoring the bending of corresponding normal and shear springs (with respective spring constants $k_n$ = 150 N/m and $k_s$ = 300 N/m), via FECO and an air-gap capacitor respectively. Shear forces were measured while applying lateral back-and-forth motion to the top surface via a sectored piezoelectric tube (PZT). Force profiles are normalized as ($F_n/R$) vs. $D$ in the Derjaguin approximation (which is valid here since $R \gg D$ and yields the interaction-energy/area between flat parallel surfaces obeying the same force law [43]).

The separation distance $D$ presented is with respect to the pre-calibrated mica-mica contact separation. After calibrating mica - mica contact in air, a droplet (0.15 mL) of hSF-PLs liposome dispersion was injected between the two mica surfaces, separated by ca. 2 mm, forming a meniscus, and left to incubate for at least 2 hours. Water was added to the boat to keep a constant humidity which prevented water evaporation from the meniscus, and force profiles then measured, following which calcium nitrate was added to the inter-surface meniscus to get a final concentration of calcium at ca. 2 mM. This compares with free calcium concentrations in the range 1.5 – 4.7 mM for healthy hSF [[44], [45]], and 2.1±0.3 mM for OA hSF [46]. Force profiles were taken following overnight incubation. Temperature of the system was maintained constant at 25 ± 1 ºC. The surfaces did not pass air/water interface throughout the SFB measurements. Data is from two completely independent experiments (i.e different pairs of mica substrates) using one batch of liposome dispersion with a number of independent contact points in each. Since evaluation of the force profiles at each of the independent contact points may be viewed as a separate determination of the surface interactions (both normal and shear), the general agreement (within the scatter) between these several independent measurements allows us to have confidence in the generalizability of the results.

## 3. Results and Discussion

### 3.1 Characterization of hSF-PLs-SUVs in bulk and on mica

The composition of PLs extracted from 17 hSF samples at a similar OA stage was characterized earlier [[35], [36]] and is presented in Fig. 2. Compared with healthy hSF (206 nmol/mL), OA-hSF of this group has a 2.6-fold higher concentration of total PLs (546 nmols/mL; $P = 0.0035$) [35]. Shown superposed on these mean values in Fig. 2 are the corresponding values of the PLs used in our study, taken from a patient at a similar OA stage. As seen, they are almost entirely within the spread of mean values for the 17 hSF samples. The PL classes identified in OA-hSF at this stage include PC, lyso PC (LPC), ether-phosphatidylcholine (PC O), phosphatidylethanolamine (PE), phosphatidylethanolamine-based plasmalogen (PE p), sphingomyelin (SM), ceramide (Cer), and negatively charged PLs such as phosphatidylglycerol (PG) and phosphatidylinositol (PI).

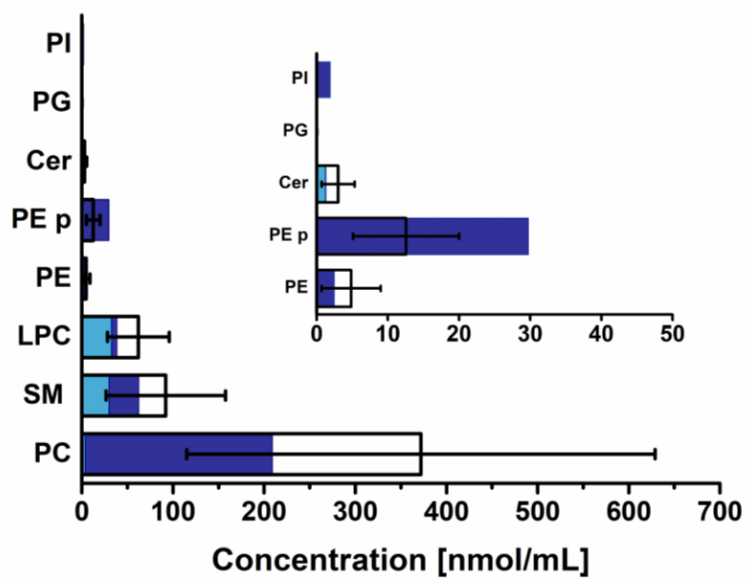

**Fig. 2.** Mean values of concentrations of major PL classes identified in hSF of patients with early-OA ($n$ =17; data are based on Ref. [35]) shown as white columns, with error bars indicating corresponding standard deviation values. The corresponding values for the sample used in this study are superposed as dark (unsaturated) and light (saturated) blue columns, showing that (with the exception of the low concentration component PE p and negatively charged PLs) they are all within the standard deviation for this OA stage. Inset shows low-concentration PLs.

The size distribution of the prepared vesicles from the OA-hSF sample is shown in Fig. 3. DLS measurements show that the average peak diameter for prepared vesicles are 128 nm and the polydispersity index (PDI) value is 0.30, indicating that the vesicles' size distribution is relatively wide. Results obtained from NTA (which samples a different size-average) show that the average size of vesicles is 101 ± 30 nm, and the approximate particle concentration is (2.7 ± 1.0) × $10^{10}$ particles/mL. Smaller particle size and narrower distribution obtained by NTA than DLS has also been indicated elsewhere [47]. This may be attributed to the sensitivity of the DLS method to larger particle sizes, as well as to dilution with water (in the DLS measurements) which might cause the liposomes – originally prepared in salt solution – to swell. Zeta potential values for vesicles prepared from the OA-hSF (diluted by 20-fold with water to 7.5 mM NaNO$_3$) are found to be -18.6 ± 6.7 mV. The negative charge is attributed primarily to anionic PLs such as PA and PG [[35], [36]], as well as zwitterionic PE which shows some anionic characteristics [48]. The presence of PPIs, that are mostly positively charged under neutral pH in the SUV dispersions, did not change the overall negative charge characteristics of the vesicles, possibly because their total concentration is very low compared with other lipids composing the vesicles.

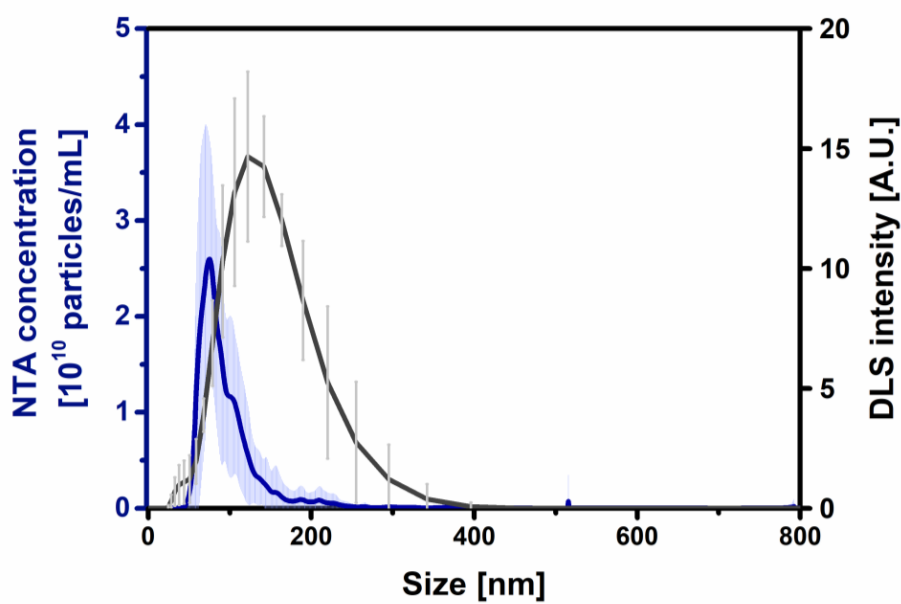

**Fig. 3**. Size distribution measured with NTA (blue) and DLS (black) methods. The DLS results represent size distribution by intensity. The NTA test allows visual, particle-by-particle validation of the data and gives approximate particle concentration value. The liposome sample prepared in 150 mM NaNO$_3$ was diluted by 50-fold with the same salt solution for NTA and by 20-fold with water for DLS. Light blue shaded area and light gray bars represent the standard deviation around the mean value.

The interaction between PL membranes that consist of zwitterionic PLs alone is significantly different from those mixed with anionic ones [[49], [50]]. The negative charge introduces repulsive electrostatic interactions between PL membranes and with other biomolecules of the same charge [[51], [52]]. Since the biomacromolecules in synovial joints covering the cartilage surface (such as HA and lubricin) are negatively charged, multi-valent cations, therefore, can act as binding agents for PL bilayers to these biomacromolecules [41]. In addition, divalent cations such as calcium and magnesium that are present in hSF induce phase-separated regions in anionic-zwitterionic PLs membranes and fusion of vesicles containing anionic PLs [[49], [53]].

The AFM scans demonstrate that when the hSF-PLs vesicles are adsorbed on mica they rupture and form a continuous bilayer with some small defects (Fig. 4(a)). A similar picture was observed after calcium ions were added to the bilayer (Fig. 4(b)). This is possibly due to the fact that the majority of PLs in hSF are unsaturated, with low main transition temperatures, and are in the fluid state under room temperature, and is in line with the behaviour of single-component fluid-phase PCs, which may readily heal in the presence of free lipids in the surrounding liposomal dispersion [[25], [54]]. Despite negative charges on the vesicle surface, their adsorption onto negatively-charged mica was made possible through screening the vesicle-mica charge-charge repulsion by high concentration of monovalent salt, and/or by bridging with divalent cations [[55], [56]]. We note from Fig. 4(c) that similar bilayers form on the mica surfaces also in the SFB, though the passage through the air/water interface, necessary for the AFM imaging of these surfaces, damages the bilayer, forming holes as shown. There is also some indication for lipid phase separation in Fig. 4(c), manifested as lighter (slightly higher) patches in the micrograph.

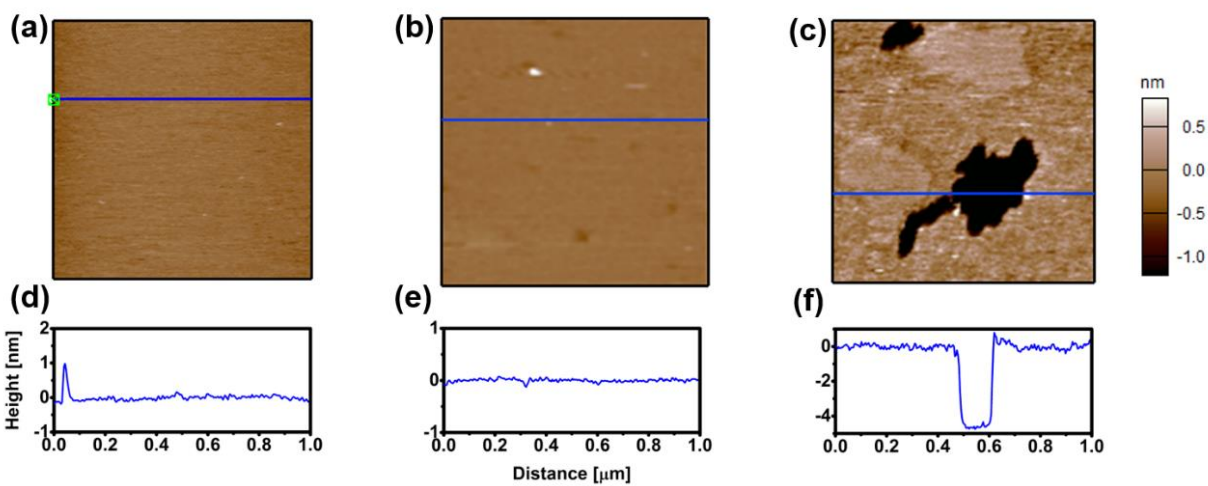

**Fig. 4**. AFM images under 150 mM $NaNO_3$ of freshly cleaved mica immersed in 0.3 mM PLs vesicle dispersion (prepared in 150 mM $NaNO_3$). (a) and (d): prior to adding calcium. (b) and (e): after adding ca. 2 mM $Ca(NO_3)_2$. (c) and (f): surface of the lens-mounted mica after SFB experiments. Image (c) was obtained after an SFB experiment across 150 mM $NaNO_3$, and the lens passed air/water interfaces twice when transferring from the SFB boat to the AFM sample holder. As the majority of PLs in OA-hSF are unsaturated with phase transition temperatures lower than room temperature, the bilayer formed on mica is fluid and cannot maintain its integrity without PLs in the surrounding medium. Phase separation is indicated as lighter patches in the PL bilayer in (c).

### 3.2 SFB measurements: Normal Forces

Normalized normal force versus separation distance ($F_n/R$ vs. $D$) profiles between two mica surfaces across ca. 0.3 mM liposome dispersion in 150 mM $NaNO_3$ before and after adding calcium are present in Fig. 5(b). A weak repulsive force commences at $60 \pm 20$ nm, attributed to steric forces of removing excess lipids from the contact region, as seen in other SFB studies of lipid-coated surfaces [30]. The repulsion monotonously increases to a 'hard-wall' at $D = 5.5 \pm 1.2$ nm and $5.3 \pm 1.1$ nm before and after adding calcium respectively. As indicated by AFM (Fig. 4(c)), the height of lipid patches observed after the lens passed air/water interfaces and scanned in 150 mM $NaNO_3$, without self-healing materials in the environment, is ca. 5 nm, corresponding to the thickness of a fluid-phase PL bilayer [57].

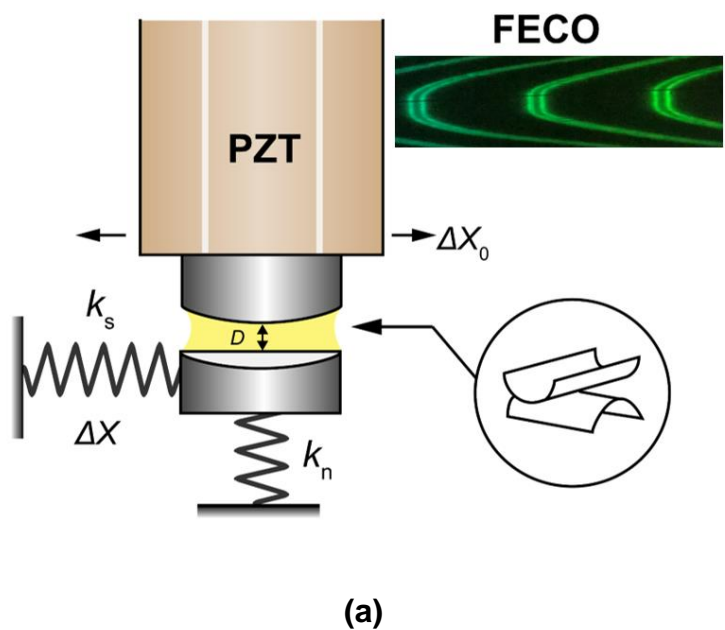

(a)

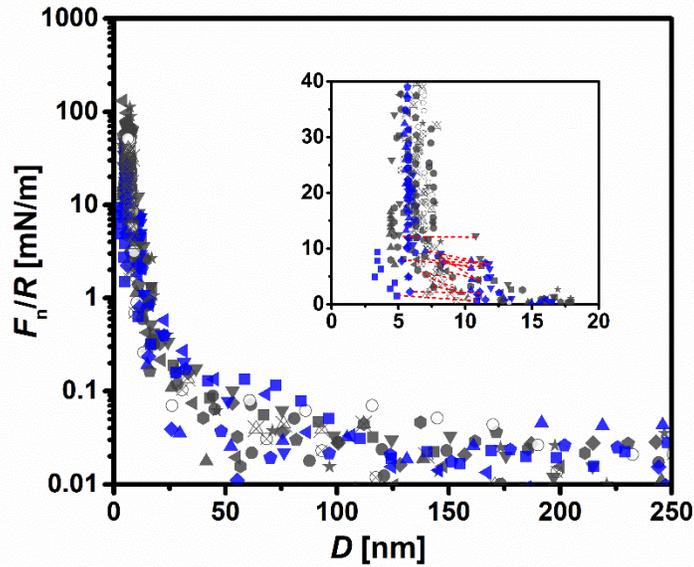

**(b)**

**Fig. 5**. A schematic of the SFB (a) and normalized normal force versus separation distance ($F_n/R$ vs. $D$) profiles between two mica surfaces immersed in ca. 0.3 mM vesicles prepared from PLs mixture extracted from OA-hSF 150 mM NaNO$_3$, before (gray data points) and after (blue) adding ca. 2 mM Ca(NO$_3$)$_2$ (b). In figure (a), the separation ($D$) between the two surfaces was measured from the wavelengths of FECO obtained by multiple beam interference. Shear force was measured by applying back-and-forth motions $\Delta x_0$ via the fsectored PZT holding the upper lens. Normal and shear forces were calculated according to the bending of corresponding springs $k_n$ and $k_s$, measured respectively via $D$ and via an air-gap capacitor. Experiments were carried out by adding a droplet of liposome dispersion (yellow in figure (a)) to the gap between the back-silvered curved mica surfaces, forming a meniscus. In figure (b), the filled and empty symbols represent the first and second approaches, and the crossed ones represent data for receding profiles. Broken red lines in the inset of figure (b) indicate inward motion to hemifusion.

The normal force profiles may be understood as follows: the mica surfaces immersed in the liposomal dispersion are each covered with an identical PL bilayer (see Fig. 4), and on compression these undergo hemifusion to reach the 'hard-wall' separation $D \approx 5$ nm which is the thickness of a single bilayer. Prior to hemifusion, the weak repulsive forces observed between two PL bilayer arise from repulsive undulations, electrostatic repulsion as well as hydration repulsion, overwhelming the attractive van der Waal's force [[58], [59]]. Following hemifusion, the repulsive force rises steeply as the separation distance decreases by a few angstroms when higher normal load compresses the confined bilayer.

### 3.3 SFB measurements: Shear forces

Frictional forces $F_s$ were measured by recording directly the shear-force traces as the surfaces move laterally past each other at different compressive loads $F_n$; typical traces are shown in the inset to Fig. 6(b). We note a small jump-in was observed in some approach profiles from ca. 11 to 6 nm, that was accompanied by an abrupt increase in the friction forces (Fig. 6(a)), indicating a hemifusion process, as was observed previously [32]. The normal force measured just before the hemifusion, is 8.6 ± 3.2 mN/m, equivalent to (1.4 ± 0.5) MPa from the Hertzian expression $P = F_n/(\pi*(F_n*R/K)^{2/3})$, where $F_n$ is the normal force, the effective mica/glue modulus $K = (1 \pm 0.3) \times 10^9$ N/m$^2$ [43], and $R \approx 10^{-2}$ m is the mean radius of curvature of the mica sheets [60]. Interestingly, this value is significantly lower than the critical pressure 5.0 ± 0.5 MPa previously reported for hemifusion of bilayers of a DPPC-POPC (2:8, molar ratio) mixture [32]. Since hemifusion is triggered by long-ranged hydrophobic force between exposed hydrophobic tails and the energy barrier for hemifusion between PL bilayers is lowered by the mismatch of the PLs forming the bilayers [61], the more heterogenous composition of PLs in hSF likely facilitates the hemifusion between the corresponding bilayers. Moreover, the presence of triacylglycerols, confirmed by LC-MS (Fig. S1), and PPIs in the PLs mixture also promotes hemifusion between apposing PL bilayers.

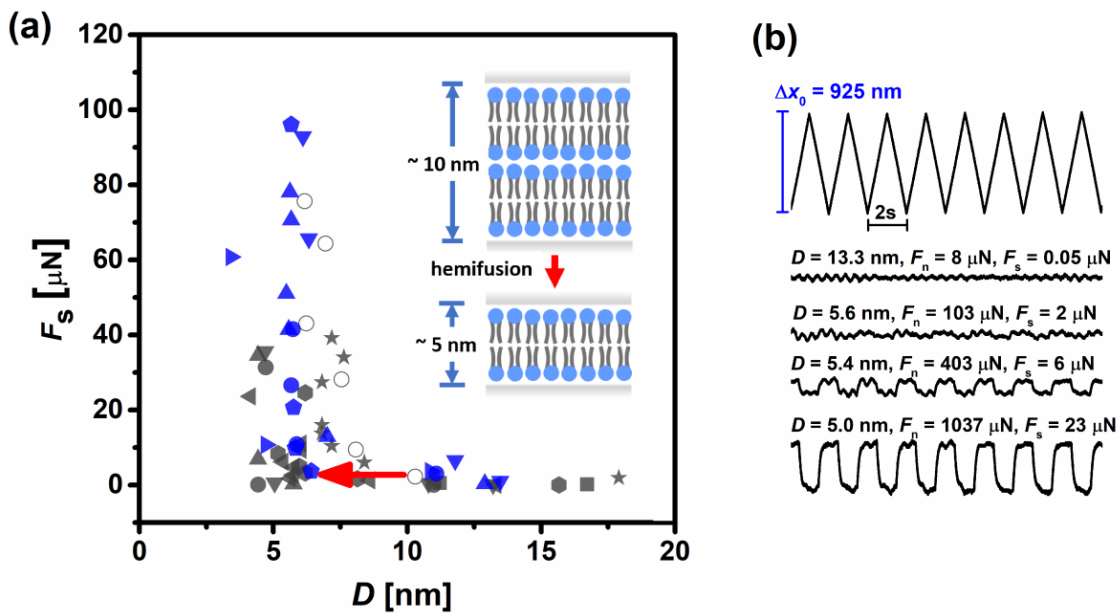

**Fig. 6**. (a) Shear force versus separation distance ($F_s$ vs. $D$) profiles between mica surfaces immersed in ca. 0.3 mM vesicles prepared from PLs extracted from OA-hSF in 150 mM NaNO$_3$ before (gray) and after (blue) adding ca. 2 mM Ca(NO$_3$)$_2$. Filled and empty symbols represent the first and second approaches respectively. The arrow indicates hemifusion. (b) The applied back-and-forth motion via PZT (most upper trace) and typical responsive shear force versus time traces taken at different separation distances and normal loads from the same force profile prior to adding calcium. The cartoon in (a) shows a schematic illustration of the hemifusion process, where two PL bilayers hemi-fused into one, accompanied by a ca. 5 nm decrease in $D$.

In Fig. 6(a), shear force versus separation distance ($F_s$ vs. $D$) profiles show that the frictional force is low when two mica surfaces are separated by more than 10 nm, corresponding to at least two bilayers trapped between the mica surfaces. In that case, the slip plane is assumed to be located at the highly hydrated headgroup/headgroup interface. The low friction due to hydration lubrication takes place at the hydrated PL headgroup layers even at physiologically high salt concentrations [[30], [56]]. When the separation distance decreases to 6.1 ± 1.2 nm at a relatively low normal load, the sliding friction rises, which we attribute to a shift of the slip plane from a midplane between two bilayers (when $D \gtrsim 10 - 11$ nm) to another interface, as discussed below. On further compression the shear force increases rapidly with a slight change in the separation distance (0.8 nm) when higher normal load is applied to the system (Figs. 5 – 7). Two frictional regimes may be identified, prior to and following the addition of calcium ions to the aqueous liposome dispersion. As can be seen in Fig. 7, a linear relation between shear and normal forces was observed for both, with friction coefficients $\mu = 0.03 \pm 0.004$ and $\mu = 0.2 \pm 0.02$ before and after adding calcium respectively. Thus adding calcium to the system, while it does not change the normal force profiles significantly (Fig. 5(b)) (with hemifusion occurring with or without calcium addition), nonetheless substantially increases the friction coefficients after hemifusion (Fig. 7). Indeed, calcium addition increases the friction by close to an order of magnitude.

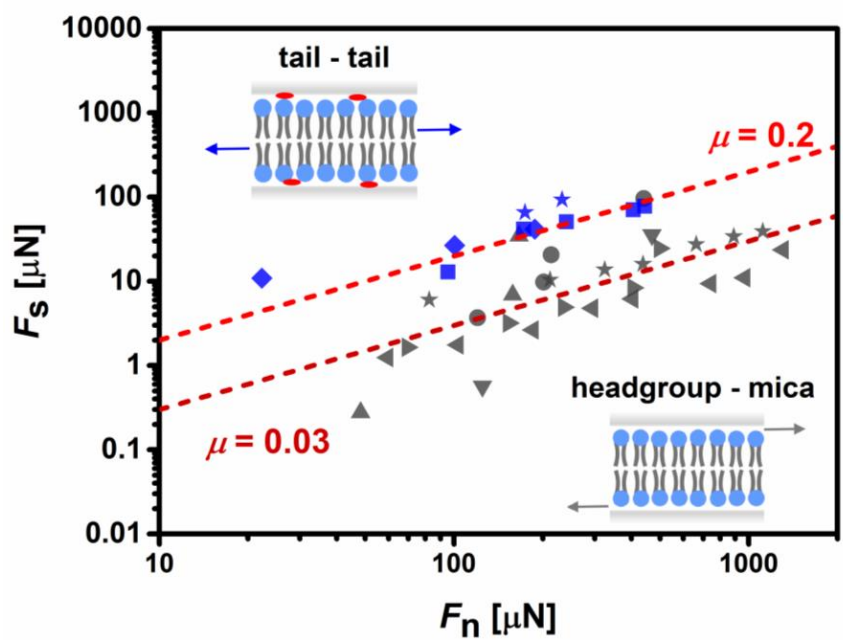

**Fig. 7**. Shear force versus normal force ($F_s$ vs. $F_n$) profiles between mica surfaces immersed in ca. 0.3 mM vesicles prepared from PLs extracted from OA-hSF in 150 mM $NaNO_3$ before (gray) and after (blue) adding ca. 2 mM $Ca(NO_3)_2$. All the data points were obtained in the compressed bilayer state, at $D < 10$ nm. Dashed lines indicating friction coefficients of 0.2 (upper) and 0.03 (lower) are guides for eyes. Insets are schematic illustrations showing slip planes in hydrophobic-tail/hydrophobic-tail (top) and lipid-headgroup/mica (bottom) interface under conditions with and without added calcium.

We attribute this behaviour as follows: following hemifusion from two bilayers to one (inset in Fig. 6(a)), the mid-plane interface between two bilayers is eliminated, and the slip plane must shift either to the hydrophobic-tails/hydrophobic-tails interface, or to the lipid-headgroup/mica interface. It is instructive to consider friction coefficients determined independently for these two types of interfaces. For sliding in air between two smooth surfactant monolayers, each exposing a layer of alkyl chains, friction coefficients $\mu \approx 0.05 - 0.1$ have been reported [62], while for interdigitating alkyl chains sliding past each other in air and water, values $\mu \approx 0.5 \pm 0.05$ have been measured [63], likely due to the greater energy dissipation on sliding the interdigitated chains past each other. From these values we may infer that in the absence of calcium, the slip plane is likely to be that of the hydrated headgroups of the bilayer in contact with each of the mica surfaces (bottom right inset to Fig. 7). Sliding there may occur relatively easily, helped by the hydration lubrication mechanism, corresponding to the observed $\mu = 0.03$. This value of $\mu$ is higher than for friction between two hydrated headgroup layers [[12], [30]], as there is likely to be an attractive component due to dipole-charge attraction between any zwitterionic lipids in the bilayer and the negatively-charged mica. Independent indication for this is provided in a recent investigation of lubrication by bilayers consisting of lipid mixtures (unpublished data).

Once calcium is added, however, the divalent calcium ions may not only bridge between negatively charged PL molecules within the mixed bilayer, but can also form bridges between the negatively charged PL bilayer and the negatively-charged mica (top left cartoon inset to Fig. 7). Overcoming such lipid-mica bridging during sliding entails considerable energy dissipation, and thus greatly increases the sliding friction. In that case we may attribute the higher value of $\mu \approx 0.2$ either to this increased dissipation on sliding at the lipid-headgroup/mica interface, or to a shift of the slip plane from the lipid-headgroup/mica to the hydrophobic-tail/hydrophobic-tail interface, where some interdigitation is expected, so that $\mu$ is likely to be in the range $0.1 - 0.5$ as noted above, consistent with the observed value $\mu \approx 0.2$. Similar results have been reported for a double-chained surfactant layer on mica [64]. These results further confirm that divalent cations play a crucial role in attaching PLs bilayers containing anionic PLs to the negatively charged substrate. In the context of synovial joint lubrication, calcium may bind negatively charged PLs layers to the cartilage surface or to the negatively-charged macromolecules (such as HA) at its outer surface. This also provides additional direct indication that in healthy synovial joints, lubrication is unlikely to take place between hydrophobic-tail/hydrophobic-tail interface under biological conditions as was earlier suggested [14].

Our findings indicate that the bilayers composed of PLs and other molecules extracted from OA-hSF, which exhibit a large lipid heterogeneity, may undergo hemifusion under pressures comparable or lower than those in joints, thereby reducing the efficacy of hydration lubrication, leading to higher friction and thus further promoting OA. The presence of PPIs and triacylglycerols (TAGs) may, by acting as 'contaminants', further lower the energy barrier for hemifusion. At the same time, a higher content of unsaturated PLs in the OA joints can create more fluid bilayers which are more likely to heal when disrupted [32]. We should note however that the composition of lipids in hSF (as used in our experiments) might be different from those in the cartilage surface boundary layer [[35], [65]]. Finally, we might inquire concerning the difference between lipids from the OA-hSF used in our study and lipids from healthy hSF (to be examined in future work). The latter too would be expected to be a lipid mixture, which would affect its surface interaction and lubrication behaviour, including hemi-fusion, relative to single component lipids. However, since the detailed behaviour must depend on the precise composition and concentration of the lipid mixture, and since that is known to differ between the OA-hSF used in our study and healthy hSF [[35], [36]], we might expect the lubrication behaviour also to differ between the two cases.

## 4. Conclusions

The mixture of PLs extracted from OA-hSF is composed of a considerable number of both zwitterionic and negatively charged PLs, so that vesicles prepared from the PLs mixture are net-negatively charged. Repulsive forces were observed between these PLs bilayers, in physiologically high salt concentrations with and without calcium, a biologically important divalent cation existing in synovial fluid. This study implies the importance of calcium in stabilizing negatively charged PL membrane to the substrate of the same charge, resembling conditions in synovial joints. Hemifusion between OA-PLs bilayers takes place at much lower normal loads than those in pressure-bearing synovial joints, resulting in the reduction of lubricious headgroup/headgroup interfaces and thus a higher overall friction, which may promote the progress of OA.

## Abbreviations

hSF – human synovial fluid, OA – osteoarthritis, PL – phospholipid, PC – phosphatidylcholine, LPC – lysophosphatidylcholine, PC O – ether-phosphatidylcholine, PE – phosphatidylethanolamine, PE p – phosphatidylethanolamine-based plasmalogen, SM – sphingomyelin, PG – phosphatidylglycerol, phosphatidylinositol (PI), Cer – ceramide, TAG – triacylglycerol, PPI – Proteinase- and phospholipase-inhibitor, SFB – surface force balance, FECO – fringes of equal chromatic order.

## Acknowledgements

This work was made possible in part by the historic generosity of the Harold Perlman family. The authors wish to express their gratitude to C. Hild for excellent technical support and the surgical team of the Department of Orthopaedics (University Hospital Giessen and Marburg GmbH, Giessen, Germany) for sampling the OA synovial fluid.

**Supplementary Material**

Liquid chromatography-mass spectrometry (LC-MS) chromatogram of lipids extracted from hSF.

**Funding sources**

Funding: This work was funded by the European Research Council [Advanced Grant CartiLube, grant number 743016], the Israel Ministry of Science and Technology [grant number 86341], the McCutchen Foundation, and the Israel Science Foundation − National Natural Science Foundation of China joint research program [grant number 2577/17].

**Conflict of Interest**

The authors declare no conflict of interest.

**CRediT author statement**

**Yifeng Cao**: Conceptualization, Methodology, Data Curation, Writing – Original Draft

**Nir Kampf**: Data Curation, Writing – Review & Editing

**Marta Krystyna Kosinska**: Data Curation, Resources

**Juergen Steinmeyer**: Resources, Writing – Review & Editing

**Jacob Klein**: Conceptualization, Writing – Review & Editing, Supervision, Project administration, Funding acquisition